\documentclass[final,authoryear,3p,times]{elsarticle}

\usepackage{graphicx,amssymb,amsmath,amsthm,bm,accents,color} % hyperref

\DeclareMathOperator{\tr}{tr}
\DeclareMathOperator{\divr}{div}
\DeclareMathOperator{\grad}{grad}
\DeclareMathOperator{\erfc}{erfc}

\newtheorem{sgrep}{Second-grade solution representation}

\newtheorem{obrep}{Oldroyd-B solution representation}

\newtheorem{theorem}{Theorem}

\journal{Mechanics Research Communications}

\begin{document}

\begin{frontmatter}

\title{Stokes' first problem for some non-Newtonian fluids: Results and mistakes}

\author{Ivan C. Christov}

\ead{christov@alum.mit.edu}
\ead[url]{http://alum.mit.edu/www/christov}

\address{Department of Engineering Sciences and Applied Mathematics, Northwestern University, Evanston, IL 60208-3125, USA}

%Tel.: +1-847-491-3555 Fax: +1-847-491-2178

\begin{abstract}
The well-known problem of unidirectional plane flow of a fluid in a half-space due to the impulsive motion of the plate it rests upon is discussed in the context of the second-grade and the Oldroyd-B non-Newtonian fluids. The governing equations are derived from the conservation laws of mass and momentum and three correct known representations of their exact solutions given. Common mistakes made in the literature are identified. Simple numerical schemes that corroborate the analytical solutions are constructed.
\end{abstract}

\begin{keyword}
Stokes' first problem \sep Second-grade fluid  \sep Oldroyd-B fluid \sep Integral transform methods \sep Finite-difference scheme
\end{keyword}

\end{frontmatter}

%\linenumbers

%% main text
\section{Introduction}
\label{sec:intro}

Following \citet{TR00}, consider a fluid of density $\varrho(\bm{x},t)$ and velocity field $\bm{u}(\bm{x},t)$, where $\bm{x}$ is the spatial coordinate and $t$ the temporal one. Conservation of mass dictates that
\begin{equation}
\dot{\varrho} + \varrho\divr\bm{u} = 0,
\label{eq:mass_cons}
\end{equation}
where a superimposed dot denotes the material derivative: $\dot{\bm{\mathfrak{F}}} = \partial\bm{\mathfrak{F}}/\partial t + (\bm{u}\cdot\grad)\bm{\mathfrak{F}}$. Cauchy's first law of continuum mechanics supplies the additional conservation of momentum equation:
\begin{equation}
\varrho\dot{\bm{u}} = \divr\mathbf{T} + \bm{b},
\label{eq:cauchy_thm}
\end{equation}
where $\mathbf{T}(\bm{x},t)$ is the stress tensor, and $\bm{b}(\bm{x},t)$ represents the body force(s).

We suppose that the fluid is incompressible and homogeneous so that $\dot{\varrho} = \grad\varrho = 0\Rightarrow \varrho = const. =: \varrho_0(>0)$ for all $\bm{x}$ and $t$. Then, Eq.~\eqref{eq:mass_cons} implies that $\divr\bm{u} = 0$, meaning that such a fluid can only undergo isochoric motions. Assuming a Cartesian coordinate system, where $\bm{x} = x\hat{\bm{\imath}} + y\hat{\bm{\jmath}} + z\hat{\bm{k}}$ with unit vectors in the three coordinate directions $\hat{\bm{\imath}}$, $\hat{\bm{\jmath}}$ and $\hat{\bm{k}}$, one such motion is the unidirectional plane flow $\bm{u} = u(y,t)\hat{\bm{\imath}}$, which clearly satisfies $\divr\bm{u} = 0$. Finally, we assume no external forces act on the fluid: $\bm{b}=\bm{0}$. All this means that the fluid fills the half-space $y>0$ with a solid plate lying in the $x$-$z$ plane (i.e., at $y=0$). The motion is uniform (translation-invariant) in the $x$ and $z$ directions.
%, is
%\begin{equation}
%[\mathbf{D}] = \left\|\begin{matrix} 0 & \frac{1}{2}\frac{\partial u}{\partial y} & 0\\ \frac{1}{2}\frac{\partial u}{\partial y} & 0 & 0\\ 0 & 0 & 0 \end{matrix}\right\|,
%\end{equation}

In 1851, Stokes considered a specific case of such a unidirectional plane flow.\footnote{Some authors misattribute this problem to Rayleigh though Stokes was the first to solve it \citep[see, e.g.,][Chap.~V \S4]{s79}.} He was interested in the case wherein the plate at $y=0$ is set into motion \emph{suddenly} at time $t=0^+$. In other words, the plate's velocity is given by $U(t) = \tilde U (t) H(t)$, where $H(\cdot)$ denotes the Heaviside unit step function, and $\tilde U(t)$ is some smooth function, i.e., it possesses as many continuous derivatives with respect to $t$ on $(-\infty,+\infty)$ as needed, that we are free to specify. Stokes himself made the distinction between $U(t)$ and $\tilde U(t)$ clear \citep[p.~101]{S51}, yet a bewildering array of papers from the 1990s and 2000s fail to take this into account. Thanks to the no-slip boundary condition, the fluid near the boundary assumes the velocity of the the plate, i.e., $u(0,t) = \tilde U(t)H(t)$. Consequently, if one needs to compute the acceleration of the fluid at the plate (e.g., when applying the Fourier sine transform to a mixed derivative), the correct expression is $\frac{\partial u}{\partial t}(0,t) = \tilde U'(t)H(t) + \tilde U(t)\delta(t)$, where $\delta(\cdot)$ denotes the Dirac delta function. Of course, such equalities are meant in the \emph{sense of distributions} as the \emph{generalized functions} $H(t)$ and $\delta(t)$ fail to have point-values everywhere \citep[\S21]{KF75}. An ubiquitous but inexcusable mistake is to drop the $\tilde U(t)\delta(t)$ term.

\subsection{Second-grade fluid}
\label{sec:sg_deriv}
The \emph{Rivlin--Ericksen fluids}, also known as \emph{fluids of grade $n$} \citep[Chap.~6]{TR00}, are a model of isotropic simple fluids of the differential type. Their constitutive relation can be written as an expansion in terms of the Rivlin--Ericksen tensors $\mathbf{A}_k$. For the incompressible second-grade fluid  it takes the form
\begin{equation}
\mathbf{T} = -p\mathbf{I} + \mathbf{S},\qquad \mathbf{S} = \mu_0\mathbf{A}_1 + \alpha_1\mathbf{A}_2 + \alpha_2\mathbf{A}_1^2,\qquad \tr\mathbf{D}=0,
\end{equation}
where $p$ is the isotropic (indeterminate) stress, $\mathbf{I}$ is the identity tensor, $\mathbf{S}$ is the extra (determinate) stress, $\mathbf{A}_1 = 2\mathbf{D}$, $\mathbf{A}_{k+1} = \dot{\mathbf{A}}_k + \mathbf{A}_k\grad\bm{u} + (\grad\bm{u})^\top\mathbf{A}_k$ ($k\ge1$),  $\mathbf{D} \equiv \tfrac{1}{2}\left[\grad\bm{u} + (\grad\bm{u})^\top \right]$ is the symmetric part of the velocity gradient (sometimes referred to as the infinitesimal rate of strain) tensor, and a $\top$ superscript denotes the transpose. The constant $\mu_0(>0)$ is understood in the usual sense of fluid viscosity from Navier--Stokes theory, and the second-grade (constant) parameters $\alpha_1$ and $\alpha_2$ will be discussed shortly. 

\citet{T63} was the first to consider plane flows of the second-grade fluid and give \emph{correct} solutions using the Laplace transform and the Bromwich integral inversion formula. However, he did not consider Stokes' first problem. Following his derivation, for which he is rarely given credit, the stress tensor for our unidirectional plane flow has the component representation
%$\bm{u}\cdot\grad\bm{A}_1 =0$, so
%\begin{equation}
%[\mathbf{A}_1] = \left\|\begin{matrix} 0 & \frac{\partial u}{\partial y} & 0\\ \frac{\partial u}{\partial y} & 0 & 0\\ 0 & 0 & 0 \end{matrix}\right\|,\qquad 
%[\mathbf{A}_1^2] = \left\|\begin{matrix} \left(\frac{\partial u}{\partial y}\right)^2 & 0 & 0\\ 0 & \left(\frac{\partial u}{\partial y}\right)^2 & 0\\ 0 & 0 & 0 \end{matrix}\right\|,\qquad
%[\mathbf{A}_2] = \left\|\begin{matrix} 0 & \frac{\partial^2 u}{\partial t\partial y} & 0\\ \frac{\partial^2 u}{\partial t\partial y} & 2\left(\frac{\partial u}{\partial y}\right)^2 & 0\\ 0 & 0 & 0 \end{matrix}\right\|, 
%\qquad  [\mathbf{A}_2^2] = \left\|\begin{matrix} \left(\frac{\partial^2 u}{\partial t\partial y}\right)^2 & 2\left(\frac{\partial u}{\partial y}\right)^2\frac{\partial^2 u}{\partial t\partial y} & 0\\ 2\left(\frac{\partial u}{\partial y}\right)^2\frac{\partial^2 u}{\partial t\partial y} & 2\left(\frac{\partial u}{\partial y}\right)^2 & 0\\ 0 & 0 & 0 \end{matrix}\right\|
%\end{equation}
%Consequently,
\begin{equation}
%\begin{multline}
[\mathbf{T}] = \left\|\begin{matrix} -p + \alpha_2 \left(\frac{\partial u}{\partial y}\right)^2& \mu_0\frac{\partial u}{\partial y} + \alpha_1\frac{\partial^2 u}{\partial t\partial y}& 0\\ \mu_0\frac{\partial u}{\partial y} + \alpha_1\frac{\partial^2 u}{\partial t\partial y}  & -p + (\alpha_1+2\alpha_2)\left(\frac{\partial u}{\partial y}\right)^2 & 0\\ 0 & 0 & -p \end{matrix}\right\|.
%\\ 
%\Rightarrow\quad \divr\mathbf{T} = \left\{-\frac{\partial p}{\partial x} + \mu_0\frac{\partial^2 u}{\partial y^2} + \alpha_1\frac{\partial^3v}{\partial y\partial t\partial y}\right\}\hat{\bm{\imath}} + \left[-\frac{\partial p}{\partial y} + 2(\alpha_1+2\alpha_2)\frac{\partial u}{\partial y}\frac{\partial^2v}{\partial y^2}\right]\hat{\bm{\jmath}} - \frac{\partial p}{\partial z}\hat{\bm{k}}.
\label{eq:stress_sg}
\end{equation}
%\end{multline}
Substituting Eq.~\eqref{eq:stress_sg} into Eq.~\eqref{eq:cauchy_thm}, we obtain
\begin{equation}
\varrho_0 \frac{\partial u}{\partial t} = -\frac{\partial p}{\partial x} + \mu_0\frac{\partial^2 u}{\partial y^2} + \alpha_1\frac{\partial^3 u}{\partial y\partial t\partial y},\qquad 0 = -\frac{\partial p}{\partial y} +  2(\alpha_1+2\alpha_2) \frac{\partial u}{\partial y}\frac{\partial^2 u}{\partial y^2},\qquad 0 = -\frac{\partial p}{\partial z}.
\label{eq:sg_fluid_mom}
\end{equation}
Since the plate is infinite, translational invariance in the $x$-$z$ plane implies that the pressure cannot depend on $x$ or $z$ \citep[\S 61]{FW03}, i.e., $\partial p/\partial x = \partial p/\partial z = 0$. This leads to $p = p(y,t) = p_\infty + (\alpha_1+2\alpha_2) (\partial u/\partial y)^2 $ from the second equation in Eq.~\eqref{eq:sg_fluid_mom}, where $p_\infty$ is the ambient pressure and at most a function of $t$. The determination of the pressure is always overlooked in papers on this topic, yet it is a fundamental part of the solution to this problem.

Finally, we note the following thermodynamic restrictions: $\alpha_1\ge 0$ and $\alpha_1 + \alpha_2 = 0$ \citep{DR95}. When $\alpha_1 < 0$, the problem becomes ill-posed in the sense of Hadamard \citep{CDM65}. 

\subsection{Oldroyd-B fluid}
\label{sec:ob_deriv}
\citet{O50} proposed a number of constitutive relations for incompressible fluids with fading strain memory (retardation) exhibiting stress relaxation. The so-called incompressible \emph{Oldroyd-B} fluid is the one with
\begin{equation}
\mathbf{T} = -p\mathbf{I} + \mathbf{S},\qquad \mathbf{S} + \lambda_1\accentset{\bigtriangledown}{\mathbf{S}} = \mu_0(\mathbf{A}_1 + \lambda_2 \accentset{\bigtriangledown}{\mathbf{A}}_1),\qquad \tr\mathbf{D}=0,
\label{eq:ob_const_rel}
\end{equation}
where the \emph{upper-convected} time derivative \citep[Sec.~3(a)]{O50} is given by
\begin{equation}
\accentset{\bigtriangledown}{\bm{\mathfrak{F}}} = \dot{\bm{\mathfrak{F}}} - (\grad\bm{u})^\top\bm{\mathfrak{F}} - \bm{\mathfrak{F}}\grad\bm{u} + (\divr\bm{u})\bm{\mathfrak{F}}.
\end{equation}
Here, $\lambda_1$ and $\lambda_2$ are the \emph{relaxation time} and \emph{retardation time}, respectively, and $\mu_0$ is the fluid's viscosity (as before). Noting that $\mathbf{S} = \mathbf{S}(y,t)$ due to translation-invariance in the $x$-$z$ plane, the second equation in Eq.~\eqref{eq:ob_const_rel} has the component form: % ($\Rightarrow \bm{u}\cdot\grad\mathbf{S}=0$)
\begin{multline}
\left\|\begin{matrix} S_{xx} + \lambda_1 \frac{\partial S_{xx}}{\partial t}  - \lambda_1S_{yx} \frac{\partial u}{\partial y} -  \lambda_1S_{xy} \frac{\partial u}{\partial y} &  S_{xy} + \lambda_1 \frac{\partial S_{xy}}{\partial t} - \lambda_1 S_{yy}\frac{\partial u}{\partial y} &  S_{xz} + \lambda_1 \frac{\partial S_{xz}}{\partial t} - \lambda_1 S_{yz}\frac{\partial u}{\partial y} \\  S_{yx} + \lambda_1 \frac{\partial S_{yx}}{\partial t} - \lambda_1 S_{yy}\frac{\partial u}{\partial y} & S_{yy} + \lambda_1 \frac{\partial S_{yy}}{\partial t}&  S_{yz} + \lambda_1 \frac{\partial S_{yz}}{\partial t}\\  S_{zx} + \lambda_1 \frac{\partial S_{zx}}{\partial t} - \lambda_1S_{zy} \frac{\partial u}{\partial y}&  S_{zy} + \lambda_1 \frac{\partial S_{zy}}{\partial t} & S_{zz} + \lambda_1 \frac{\partial S_{zz}}{\partial t} \end{matrix}\right\| \\
= \left\|\begin{matrix} - 2\mu_0\lambda_2\left(\frac{\partial u}{\partial y}\right)^2 & \mu_0\frac{\partial u}{\partial y} + \mu_0 \lambda_2 \frac{\partial^2 u}{\partial t \partial y}& 0\\ \mu_0\frac{\partial u}{\partial y} + \mu_0 \lambda_2 \frac{\partial^2 u}{\partial t \partial y}& 0 & 0\\ 0 & 0 & 0 \end{matrix}\right\|.
\end{multline}
From the assumptions that prior to start-up the fluid is at rest, we have that $\mathbf{S}(y,0) \equiv \bm{0}$, whence the equations for the components $S_{xz}$, $S_{zx}$, $S_{yy}$, $S_{yz}$, $S_{zy}$ and $S_{zz}$ give 
\begin{equation}
S_{xz} = S_{zx} = S_{yy} =  S_{yz} = S_{zy} = S_{zz} \equiv 0 \quad\forall t \ge 0,\;\; y\ge0.
\label{eq:zero_comps}
\end{equation}
Since the stress tensor must be symmetric, i.e., $\mathbf{T} = \mathbf{T}^\top$ ($\Rightarrow S_{xy} = S_{yx}$, in particular), the remaining equations are
\begin{equation}
S_{xy} + \lambda_1 \frac{\partial S_{xy}}{\partial t} = \mu_0\frac{\partial u}{\partial y} + \mu_0 \lambda_2 \frac{\partial^2 u}{\partial t \partial y},\qquad S_{xx} + \lambda_1 \frac{\partial S_{xx}}{\partial t} -2 \lambda_1S_{xy} \frac{\partial u}{\partial y} = - 2\mu_0\lambda_2\left(\frac{\partial u}{\partial y}\right)^2.
\label{eq:extra_stress}
\end{equation}
This derivation was first given by \citet{T62}, though many authors employ a clumsy and abbreviated version of it without giving him any credit. \citet[Sec.~4]{O50} presents a similar derivation for a problem in polar coordinates.

Substituting the first equation in Eq.~\eqref{eq:ob_const_rel} into Eq.~\eqref{eq:cauchy_thm} and recalling that $\mathbf{S} = \mathbf{S}(y,t)$, we obtain
\begin{equation}
\varrho_0\frac{\partial u}{\partial t} = -\frac{\partial p}{\partial x} + \frac{\partial S_{xy}}{\partial y},\qquad 0 = -\frac{\partial p}{\partial y} + \frac{\partial S_{yy}}{\partial y},\qquad 0 = - \frac{\partial p}{\partial z}.
\label{eq:ob_mom_eqs}
\end{equation}
As in Sec.~\ref{sec:sg_deriv}, translation invariance in the $x$-$z$ plane implies $\partial p/\partial x = \partial p/\partial z = 0$. Then, $p = p(y,t) = p_\infty + S_{yy}(y,t) = p_\infty$ thanks to Eq.~\eqref{eq:zero_comps} and the second equation in Eq.~\eqref{eq:ob_mom_eqs}. Next, we can eliminate $S_{xy}$ from the first equation in Eqs.~\eqref{eq:extra_stress} and \eqref{eq:ob_mom_eqs} by taking the $y$ derivative of the former and the $t$ derivative of the latter. Thus, we arrive at
\begin{equation}
\varrho_0\frac{\partial u}{\partial y} + \varrho_0\lambda_1\frac{\partial^2 u}{\partial t^2} = \mu_0 \frac{\partial^2 u}{\partial y^2} + \mu_0\lambda_2\frac{\partial^3 u}{\partial y \partial t \partial y}.
\label{eq:ob_fluid_mom}
\end{equation}
The two non-trivial components $S_{xx}$ and $S_{xy}$ of the determinate stress can be calculated from Eq.~\eqref{eq:extra_stress} once $u$ is found from Eq.~\eqref{eq:ob_fluid_mom}. This completes the formulation of the problem.

Considerations from thermodynamics \citep{RS00} restrict the values of the relaxation and retardation times to be such that $\lambda_2 < \lambda_1$, though here we present some solutions also valid for $\lambda_2 \ge \lambda_1$. Causality requires that $\lambda_1>0$. Then, for the problem to be well-posed in the sense of Hadamard, a necessary (but not sufficient) condition is that $\lambda_2>0$. Experimental observations support all of these restrictions \citep{TS53,O58}.

\section{Exact solutions by integral transform methods}
\label{sec:transforms}

\subsection{Second-grade fluid}
\label{sec:sg_soln}
Defining $\nu := \mu_0/\varrho_0$ and $\alpha := \alpha_1/\varrho_0$ and supplying Eq.~\eqref{eq:sg_fluid_mom} with the boundary condition discussed in Sec.~\ref{sec:intro} and a proper decay condition as $y\to\infty$, we have the following initial-boundary-value problem (IBVP):
\begin{subequations}\label{eq:sg_ibvp}\begin{align}
\frac{\partial u}{\partial t} &= \nu \frac{\partial^2 u}{\partial y^2} + \alpha \frac{\partial^3 u}{\partial y\partial t\partial y}, &&(y,t)\in(0,\infty)\times(0,\infty);\label{eq:sg_pde}\\
u(0,t) &= U_0H(t),\qquad u \to 0 \;\;\text{as}\;\; y\to\infty, &&t>0;\label{eq:sg_bc}\\
u(y,0) &= 0, &&y>0.\label{eq:sg_ic}
\end{align}\end{subequations}

\begin{sgrep}[\citealp*{CC10}]
Assume $\alpha>0$. Using first the Fourier sine transform in $y$ and solving the resulting ordinary differential equation in $t$ with the Laplace transform, one obtains
\begin{equation}
u(y,t) = U_0H(t)\left[1-\frac{2}{\pi}\int_{0}^{\infty}\frac{\sin(\xi y)}{\xi} \exp\left(\frac{-\nu \xi^2t}{1+\alpha\xi^2}\right)\,\mathrm{d}\xi+ \frac{2\alpha}{\pi}\int_{0}^{\infty}\frac{\xi \sin(\xi y)}{1+\alpha\xi^2}\exp\left(\frac{-\nu \xi^2 t}{1+\alpha\xi^2}\right)\,\mathrm{d}\xi\right].
\label{eq:soln_cc}
\end{equation}
\end{sgrep}

While many authors have attempted to obtain this solution, \citet{CC10} show that all of them make the mistake of dropping the $\tilde{U}(t)\delta(t)$ term (recall the discussion in Sec.~\ref{sec:intro}) in the expression for the plate's acceleration when applying the Fourier sine transform. A series of papers \citep{E03,EI05,EI07a,EI07b} promulgates this error, while incorrectly claiming that there is a ``deeper'' mathematical reason that their erroneous solution does not agree with the correct Laplace transform solution. Meanwhile, three recent papers \citep{ZF07,ZBF07,ZB08} make use of the incorrect version of Eq.~\eqref{eq:soln_cc} to perform some manipulations rendering their results erroneous. The transform error is also committed in \citep{FF02,STZM06}, wherein an incorrect solution of Stokes' first problem for a second-grade fluid over a heated plate is obtained. Similarly, \citet{KAHF08} obtain erroneous solutions for MHD second-grade fluid flows.

\begin{sgrep}[\citealp*{P84}] 
Assume $\alpha > 0$. Using the Laplace transform in $t$ and the Bromwich integral inversion formula, one obtains
\begin{equation}
u(y,t)=U_0H(t)
\left[1-\frac{1}{\pi}\int_{0}^{\alpha^{-1}}\!{\rm e}^{-\eta \nu t} \sin\left( \frac{y}{\sqrt{\alpha}}\sqrt{\frac{\eta}{\alpha^{-1}-\eta}}\,\right)\!\frac{{\rm d}\eta}{\eta}\right].
\label{eq:soln_puri}
\end{equation}
\end{sgrep}

Note that Eq.~\eqref{eq:soln_cc} can be transformed into Eq.~\eqref{eq:soln_puri} by the substitution $\eta=\xi^2/(1+\alpha^2\xi^2)$. This solution representation is correctly generalized to the case of a porous half-space by \citet{JP03}, while an erroneous version of the porous-half-space problem for the related Burgers fluid is corrected in \citep{J10}.

\begin{sgrep}[\citealp*{BRG95}]
Assume $\alpha>0$. Using the Laplace transform in $t$, the quotient splitting technique of \citet{M56} and standard Laplace inversion tables, one obtains
\begin{equation}
u(y,t) = U_0H(t) \left[ {\rm e}^{-t(\nu/\alpha)}\int_0^\infty {\rm e}^{-\zeta} I_0\left(2\sqrt{\zeta t(\nu/\alpha)} \right)\erfc\left(\frac{y}{2\sqrt{\alpha}\sqrt{\zeta}} \right)\,{\rm d}\zeta \right],
\label{eq:soln_bandelli}
\end{equation}
where $\erfc(\cdot)$ is the complementary error function and $I_0(\cdot)$ the modified Bessel function of the first kind of order zero.
\end{sgrep}
Note that both \citet{BRG95} and \citet{P84} omit that the $H(t)$ pre-factor. This is not a triviality because, as the following theorem shows, $\lim_{t\to0^+} u(y,t) \ne 0 = u(y,0)$ for this problem. This is known as the \emph{start-up jump} and its physical significance is gracefully explained in \citep{JP03,J10}.

\begin{theorem}[\citealp*{BRG95}]
Let $u(y,t)$ be a function for which $\partial u/\partial y$, $\partial u/\partial t$ and $\partial^2 u/\partial t\partial y$ are all integrable at $(y,t) = (0,0)$, then $|u(\epsilon,0) - u(0,\epsilon)| \to 0$ as $\epsilon\to 0^+$.
\end{theorem}
The contrapositive of this theorem states that if $|u(\epsilon,0) - u(0,\epsilon)| \not\to 0$ as $\epsilon\to 0^+$ (i.e., the initial and boundary data are \emph{incompatible}), the solution to the IBVP may fail to have integrable either first and/or mixed second derivatives at $(y,t)=(0,0)$. This means that the solution itself is ill-behaved (singular) there. Indeed, for Stokes' first problem we have incompatible initial and boundary data. Therefore, from \citep[Eq.~(3.10)]{BRG95}, we have $|u(y,t) - u(y,0)| \not\to 0$ as $t\to0^+$, whence $\lim_{t\to0^+} u(y,t) \ne \lim_{t\to0^-} u(y,t)$.\footnote{It appears that \citet{T62} was the first one to realize this for the Oldroyd-B fluid. He notes that ``the integral form ... does not satisfy the ... conditions at $t=0$, there being no derivative at that point.'' Yet, worries about this fact prevented \citet{A69} from obtaining the solution in Eq.~\eqref{eq:soln_cc}, though he had all the ``pieces'' of it. Meanwhile, \cite{T63} shows his solutions for flows of second-grade fluids have continuous derivatives at $t=0$, however, all of these are for \emph{compatible} initial and boundary conditions.} 
This fact about limits of singular functions is \emph{completely unrelated} to integral transform methods. In the recent literature, one can find baffling statements such as ``it was shown that the previous attempts to solve the problem by using the Laplace transform technique are erroneous and that the method of the Laplace transform does not work for this problem'' \citep{E03}. However, such a statement is patently false as the two Laplace and one Fourier sine transform solutions \emph{above} are all correct, they satisfy all conditions imposed, and they are equivalent. 
%\footnote{Ironically, the Laplace transforms solution, e.g., by \citet{P84}, is correct while the Fourier sine transform solution by \citet{E03} is wrong for the reasons discussed in \citep{CC10}. An even more incoherent discussion can be found in \citep{EI07b}.}
%The fact that $\lim_{t\to0^+} u(y,t) \ne \lim_{t\to0^-} u(y,t) = 0$ is due to the singularity of the solution at the origin of $(y,t)$-plane; \emph{this is a mathematical fact about the solution completely independent from the method of solution.}

\subsection{Oldroyd-B fluid}
\label{sec:ob_soln}
Again, defining $\nu := \mu_0/\varrho_0$ and supplying Eq.~\eqref{eq:ob_fluid_mom} with the boundary condition discussed in Sec.~\ref{sec:intro} and a proper decay condition as $y\to\infty$, we have the following IBVP:
\begin{subequations}\label{eq:ob_ibvp}\begin{align}
\frac{\partial u}{\partial t} + \lambda_1 \frac{\partial u^2}{\partial t^2}  &= \nu \frac{\partial^2 u}{\partial y^2} +  \nu\lambda_2\frac{\partial^3 u}{\partial y \partial t\partial y}, && (y,t)\in (0,\infty)\times(0, \infty); \label{eq:ob_pde}\\
u(0,t) &= U_0H(t), \qquad u \to 0 \;\;\text{as}\;\; y\to\infty, &&  t>0; \label{eq:ob_bc}\\
u(y,0) &= 0,\phantom{_0H(t)}\qquad \frac{\partial u}{\partial t}(y,0) = 0, && y >0; \label{eq:ob_ic}
\end{align}\end{subequations}
%We introduce the dimensionless variables
%\begin{equation}
%t^* = t(U_0^2/\nu),\qquad y^* = y(U_0/\nu),\qquad v^* = v/U_0.
%\end{equation}
%Then, Eq.~\eqref{eq:pde2} becomes
%\begin{equation}
%\underbrace{\left(\frac{\lambda}{\nu} U_0^2\right)}_{=:\alpha}\frac{\partial^2 u^*}{\partial {t^*}^2} + \frac{\partial u^*}{\partial t^*} = \frac{\partial^2 u^*}{\partial {y^*}^2} + \underbrace{\left(\frac{\lambda_r}{\nu}U_0^2\right)}_{=:\alpha_t}\frac{\partial^3 u^*}{\partial t^*\partial {y^*}^2}.
%\label{eq:pde2-nd}
%\end{equation}
%Clearly, we can compute $v^*(y^*,t^*) = v\Big(y^*(\nu/U_0),t^*(\nu/U_0^2)
%\Big)/U_0$ given $u(y,t)$ and compare to $u(y,t)|_{\beta=0}$ from above.

\begin{obrep}[\citealp*{CJ09}]
Define $\kappa := \lambda_2/\lambda_1$ and assume $\lambda_{1,2}>0$. Using the Fourier sine transform in $y$, solving the resulting ordinary differential equation in $t$ with the Laplace transform, one obtains
\begin{equation}
u(y,t) = U_0 H(t)
\begin{cases}
1 - \tfrac{2}{\pi}\!\left[\int_{0}^{\xi_{1}^{\bullet}}\bar{\mathcal{U}}_{+}(\xi,t)\sin\left(\frac{\xi y}{\sqrt{\nu\lambda_1}}\right)\,{\rm d}\xi +\int_{\xi_{1}^{\bullet}}^{\xi_{2}^{\bullet}}\bar{\mathcal{U}}_{-}(\xi,t)\sin\left(\frac{\xi y}{\sqrt{\nu\lambda_1}}\right)\,{\rm d}\xi  + \int_{\xi_{2}^{\bullet}}^{\infty}\bar{\mathcal{U}}_{+}(\xi,t)\sin\left(\frac{\xi y}{\sqrt{\nu\lambda_1}}\right)\,{\rm d}\xi\right], &\kappa < 1,\\[2mm]
\erfc\left(\frac{y}{2\sqrt{\nu t}}\right), &\kappa = 1,\\[2mm]
1-\tfrac{2}{\pi}\!\int_{0}^{\infty}\bar{\mathcal{U}}_{+}(\xi,t)\sin\left(\frac{\xi y}{\sqrt{\nu\lambda_1}}\right)\,{\rm d}\xi, &\kappa > 1,
\end{cases}
\label{eq:soln_cj}
\end{equation}
where $\bar{\mathcal{U}}_{\pm}(\xi,t)$ correspond to $f(\xi)\gtrless 0$, respectively:
\begin{equation*}
\begin{aligned}
\bar{\mathcal{U}}_+(\xi,t)&=\frac{\exp[-g(\xi)t/\lambda_1]\left\{\sqrt{f(\xi)}\cosh[(t/\lambda_1)\sqrt{f(\xi)}]+g(\xi)\sinh[(t/\lambda_1)\sqrt{f(\xi)}]\right\}}{\xi\sqrt{f(\xi)}} -\frac{\kappa \xi \exp[-g(\xi)t/\lambda_1]\sinh[(t/\lambda_1)\sqrt{f(\xi)}]}{\sqrt{f(\xi)}},\\
\bar{\mathcal{U}}_-(\xi,t)&=
\frac{\exp[-g(\xi)t/\lambda_1]\left\{\sqrt{|f(\xi)|}\cos[(t/\lambda_1)\sqrt{|f(\xi)|}]+g(\xi)\sin[(t/\lambda_1)\sqrt{|f(\xi)|}]\right\}}{ \xi\sqrt{|f(\xi)|}} -\frac{\kappa \xi \exp[-g(\xi)t/\lambda_1]\sin[(t/\lambda_1)\sqrt{|f(\xi)|}]}{\sqrt{|f(\xi)|}},
\end{aligned}
\end{equation*}
and
\begin{equation*}
\xi_{1,2}^{\bullet} = \kappa^{-1}\sqrt{2-\kappa\mp 2\sqrt{1-\kappa}},\qquad f(\xi)=\tfrac{1}{4}\left[\kappa^2\xi^4 - 2(2-\kappa)\xi^2+1\right],\qquad g(\xi)=\tfrac{1}{2}(1+\kappa \xi^2)>0.
\end{equation*}
\end{obrep}

\begin{obrep}[\citealp*{T62}] Define $\kappa := \lambda_2/\lambda_1$ and assume $\lambda_{1,2} > 0$. Using the Laplace transform in $t$ and the Bromwich integral inversion formula, one obtains
\begin{multline}
u(y,t) = U_0H(t)\left(\frac{1}{2} + \frac{1}{\pi} \int_0^\infty \exp\left\{-\frac{y}{\sqrt{\nu\lambda_1}}\sqrt{\frac{\eta}{2}}\mathcal{M}(\eta)\big[\cos\theta(\eta) - \sin\theta(\eta) \big]\right\}\right.\\ \left.\times\sin\left\{\frac{t}{\lambda_1} \eta - \frac{y}{\sqrt{\nu\lambda_1}}\sqrt{\frac{\eta}{2}}\mathcal{M}(\eta) \big[\cos\theta(\eta) + \sin\theta(\eta) \big] \right\}\frac{{\rm d}\eta}{\eta}\right),
\label{eq:soln_tanner}
\end{multline}
where
\begin{equation*}
\mathcal{M}(\eta) = \sqrt[4]{\frac{1 + \eta^2}{1 + \kappa^2\eta^2}},\qquad \theta(\eta) = \tfrac{1}{2}\left[ \tan^{-1} \eta - \tan^{-1} (\kappa \eta)\right].
\end{equation*}
\end{obrep}

\begin{obrep}[\citealp*{M56}] Define $\kappa := \lambda_2/\lambda_1$ and assume $\lambda_{1,2} > 0$ and $\kappa \le 1$. Using the Laplace transform in $t$ and a special splitting of the resulting quotient, one can invert the transform-domain solution using standard tables of inverses to obtain
\begin{multline}
u(y,t) = U_0H(t)\exp\left[\frac{(1-\kappa)}{\kappa}\frac{t}{\lambda_1}\right] \left(\erfc\left[\frac{y}{2\sqrt{\nu t}\sqrt{\kappa}}\right] - \frac{1-\kappa}{\kappa}\int_0^{t/\lambda_1} \left\{ J_0\left[\frac{2\sqrt{1-\kappa}}{\kappa}\sqrt{\zeta \left(\frac{t}{\lambda_1}- \zeta\right)}\, \right] \right.\right.\\
+ \left.\left. \frac{\sqrt{\zeta}}{\sqrt{1-\kappa}\sqrt{t/\lambda_1 - \zeta}} J_1\left[\frac{2\sqrt{1-\kappa}}{\kappa}\sqrt{\zeta \left(\frac{t}{\lambda_1}- \zeta\right)}\, \right]\right\}  \exp\left[-\frac{(2 - \kappa)}{\kappa}\left(\frac{t}{\lambda_1} - \zeta\right)\right] \erfc\left[\frac{y}{2\sqrt{\nu \lambda_1 \zeta}\sqrt{\kappa}}\right] \,{\rm d}\zeta \right),
\label{eq:soln_morrison}
\end{multline}
where $J_p(\cdot)$ is the Bessel function of the first kind of order $p$.
\end{obrep}
\citet{M56} and \citet{T62} both neglect to multiply their solutions by $H(t)$.
%Note that the solution of Stokes' first problem for a Maxwell fluid is recovered in the limit $\kappa \to 0$ ($\lambda_2 \to 0$, $0<\lambda_1<\infty$), while the solution for a Newtonian fluid is recovered in the limit $\kappa\to 1$ ($\lambda_1=\lambda_2$ or $\lambda_{1,2}\to0$ at the same rate). Recuperating the second-grade fluid solution ($\lambda_1\to0$, $0<\lambda_2<\infty$) requires a rescaling of the independent and integration variables.
Though \citet{M56} was not studying the Oldroyd-B fluid specifically, he obtained Eq.~\eqref{eq:ob_pde} for the velocity in a viscoelastic rod whose stress response is modeled by a dashpot in series with an element consisting of another dashpot and a string in parallel.

\citet{VNFF08} claim a ``new'' solution to the IBVP in Eq.~\eqref{eq:ob_ibvp} is obtained in their paper, though they use the Laplace transform and a quotient splitting very similar to the one in \citep{M56}. Additionally, it is claimed that ``the diagrams of the solutions'' from \citep{FF03} ``are identical'' to those in \citep{VNFF08}. Though it remains unclear why a diagram (rather than an accurate plot of the solution) is relevant, the solution in \citep{VNFF08} is suspect because 
%it does \emph{not} appear to agree with Morrison's correct solution, and
it appears to agree with the \emph{incorrect} solution in  \citep{FF03}. The latter along with the solution for the porous half-space and porous half-space over a heat plate versions of this problem presented in \citep{TM05a,TM05b} were shown to be wrong by \citet{CJ09}. Other studies of Oldroyd-B \citep{F02,FHKF08,K09,FJFV09} and Burgers \citep{KMFF10} fluid flows are also erroneous because the mistake in applying the Fourier sine transform is made.

\section{Numerical solutions by finite-difference methods}
\label{sec:numerical}

To provide an independent check on the transform solutions given in Sec.~\ref{sec:transforms}, we also solve the corresponding IBVPs numerically. The (uniform) spatial and temporal step sizes are defined as $\Delta y := L/(M-1)$ and $\Delta t := t_f/(K-1)$, where $M\ge2$ and $K\ge 2$ are integers and now $(y,t)\in(0,L)\times(0,t_f]$. Also, we let $\mathfrak{u}_j^n \approx u(y_j,t^n)$ be the approximation to the exact solution on the grid, where $y_j := j\Delta y$ ($0\le j \le M-1$) and $t^n := n\Delta t$ ($0\le n \le K-1$). For appropriately chosen $L\gg1$, the front does not reach the $y=L$ boundary for any $t\in(0,t_f]$, and so this is the ``numerical infinity.''

For the computations shown below, we use $M=K=5000$ and $L = 20$ to obtain highly-accurate solutions. {\sc Matlab}'s built-in Gaussian elimination algorithm is used to invert the symmetric tridiagonal matrices resulting from the spatial discretizations. Additionally, the integral representations of the analytical solutions are evaluated using the high-precision numerical integration routine {\tt NIntegrate} of the software package {\sc Mathematica} (ver.~7.0.1).

\subsection{Second-grade fluid}
\label{sec:sg_num}
As shown in \citep{CC10}, we may discretize Eq.~\eqref{eq:sg_pde} as follows:
\begin{equation}
\delta_{t+}\mathfrak{u}^{n}_j = \nu \delta_{y+}\delta_{y-}\left[\tfrac{1}{2}\left(\mathfrak{u}^{n+1}_j + \mathfrak{u}^{n}_j\right)\right]  + \alpha \delta_{y+}\delta_{t+}\delta_{y-}\mathfrak{u}^{n}_j.
\label{eq:sg_scheme}
\end{equation}
Here, $\delta_{t+}$ is the forward temporal difference operator and $\delta_{y+}$ and $\delta_{y-}$ are, respectively, the forward and backward spatial difference operators \citep[\S3.3]{S04}. 
The boundary conditions from Eq.~\eqref{eq:sg_bc} are implemented as
\begin{equation}
\mathfrak{u}_0^n = \begin{cases}0, &n=0,\\ U_0, &1\le n\le K-1; \end{cases}\qquad \mathfrak{u}_{M-1}^n = 0,\quad 0\le n\le K-1.
\end{equation}
The initial condition is $\mathfrak{u}^0_j=0$ ($0\le j \le M-1$) owing to Eq.~\eqref{eq:sg_ic}. It is a straightforward, though lengthy, calculation \citep[see, e.g.,][]{S04} to show that this implicit, two-level Crank--Nicolson-type discretization is unconditionally stable and has truncation error $\mathcal{O}[(\Delta t)^2 + (\Delta y)^2]$. \citet{A69} gives some explicit schemes for Eq.~\eqref{eq:sg_pde}, however, the present one is superior in both its accuracy and stability.

\begin{figure}[!ht]
\centering\includegraphics[width=0.45\textwidth]{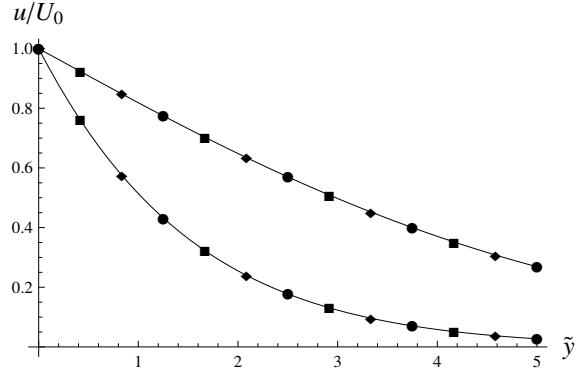}
\caption{Solutions to Stokes' first problem for the second-grade fluid for $\tilde t =1$ (bottom curve) and $\tilde t =10$ (top curve). Legend: Fourier sine transform solution from Eq.~\eqref{eq:soln_cc} ($\bullet$), Puri's Laplace transform solution from Eq.~\eqref{eq:soln_puri} ($\blacksquare$), Bandelli et al.'s  Laplace transform solution from Eq.~\eqref{eq:soln_bandelli} ($\Diamondblack$) and the numerical solution using the scheme from Eq.~\eqref{eq:sg_scheme} (--).}
\label{fig:sg}
\end{figure}

Since there exist dimensionless variables $\tilde t = t(\nu/\alpha)$ and $\tilde y = y/\sqrt{\alpha}$ such that the problem and solution no longer depend on $\nu$ and $\alpha$ \citep{CDM65,BRG95}, the qualitative shape of the solution is invariant and varying the parameters is not enlightening in any way. Therefore, in Fig.~\ref{fig:sg}, we have made use of these dimensionless variables by showing $u(\tilde y,\tilde t)/U_0$ rather than $u(y,t)$. It is clear that the three analytical solutions to the IBVP in Eq.~\eqref{eq:sg_ibvp} obtained by integral transforms presented in Sec.~\ref{sec:sg_soln} agree identically with its numerical solution. 

\subsection{Oldroyd-B fluid}
\label{sec:ob_num}
Constructing a reliable finite-difference scheme directly for Eq.~\eqref{eq:ob_pde} subject to Eq.~\eqref{eq:ob_bc} turns out to be a difficult task. An easier approach is to introduce the fluid's acceleration $w\equiv \partial u/\partial t$ with $\mathfrak{w}_j^n \approx w(y_j,t^n)$, then Eq.~\eqref{eq:ob_pde} can be trivially rewritten as a system that we discretize by the following semi-implicit Crank--Nicolson-type procedure:
\begin{equation}
\delta_{t+}\mathfrak{u}_j^{n-1/2} = \mathfrak{w}_j^n,\qquad \lambda_1\delta_{t+}\mathfrak{w}_j^{n} = -\tfrac{1}{2}(\mathfrak{w}_j^{n+1} + \mathfrak{w}_j^{n}) + \nu \delta_{y+}\delta_{y-}\mathfrak{u}_j^{n+1/2} + \nu\lambda_2\delta_{y+}\delta_{y-}\left[\tfrac{1}{2}\left(\mathfrak{w}_j^{n+1} + \mathfrak{w}_j^n\right)\right].
\label{eq:ob_scheme}
\end{equation}
Noting that $w(0,t) = U_0\delta(t)$, the boundary conditions from Eq.~\eqref{eq:ob_bc} are implemented as
\begin{equation}
\mathfrak{u}_0^{n-1/2} = \begin{cases}0, &n=0,\\ U_0, &1\le n\le K-1; \end{cases}\qquad \mathfrak{w}_0^n = \begin{cases}0, &n=0,\\ \tfrac{U_0}{n^*\Delta t}, &1\le n \le n^*,\\ 0, &n^*\le n\le K-1; \end{cases}\qquad \mathfrak{u}_{M-1}^n = \mathfrak{w}_{M-1}^n = 0,\quad 0\le n\le K-1.
\end{equation}
The initial conditions are $\mathfrak{u}^{-1/2}_j = \mathfrak{w}^0_j=0$ ($0\le j \le M-1$) owing to Eq.~\eqref{eq:ob_ic}. The scheme is not sensitive to the parameter $n^*$, so we take $n^*=10$ in our calculations.

Unfortunately, due to the implementation of the $\delta$-function boundary condition, it is no longer straightforward to show stability. Nevertheless, numerical experiments show the scheme is stable for $\Delta t = \mathcal{O}(\Delta y)$. It is easy to establish the truncation error is $\mathcal{O}[(\Delta t)^2 + (\Delta y)^2]$. \citet{T73} gives a $\mathcal{O}[(\Delta t)^2 + \Delta y]$ fully-implicit scheme for the system of Eqs.~\eqref{eq:extra_stress} and \eqref{eq:ob_mom_eqs}, however, the present scheme is simpler and more accurate. Other modern numerical approaches to one-dimensional viscoelastic flows can be found in \citep{AO10} and the references therein.

\begin{figure}[!ht]
\centering\includegraphics[width=0.45\textwidth]{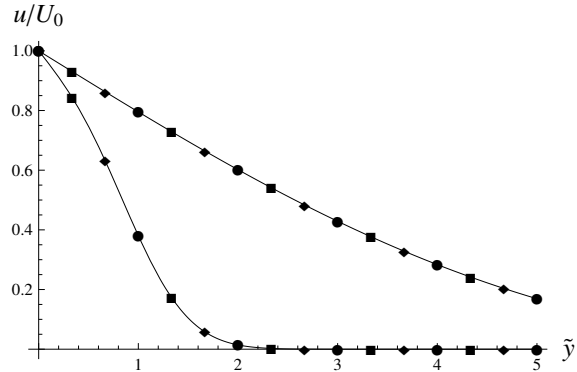}
\caption{Solutions to Stokes' first problem for the Oldroyd-B fluid with $\kappa = 0.2$ for $\tilde t =1$ (bottom curve) and $\tilde t = 7$ (top curve). Legend: Fourier sine transform solution from Eq.~\eqref{eq:soln_cj} ($\bullet$), Tanner's Laplace transform solution from Eq.~\eqref{eq:soln_tanner} ($\blacksquare$), Morrison's  Laplace transform solution from Eq.~\eqref{eq:soln_morrison} ($\Diamondblack$) and the numerical solution using the scheme from Eq.~\eqref{eq:ob_scheme} (--).}
\label{fig:ob}
\end{figure}

By using the dimensionless variables $\tilde t = t/\lambda_1$ and $\tilde y = y/\sqrt{\nu\lambda_1}$ \citep{T62}, it can be shown that qualitative differences in the shape of the solution result only from having $\kappa < 1$, $\kappa = 1$ or $\kappa > 1$. Experiments \citep{TS53} suggest that $\kappa(\equiv \lambda_2/\lambda_1)$ is in the range $0.05$ to $0.4$. Therefore, in Fig.~\ref{fig:ob}, we take $\kappa = 0.2$ and use the dimensionless variables above to show that the three analytical solutions to the IBVP in Eq.~\eqref{eq:ob_ibvp} obtained by integral transforms agree identically with its numerical solution. 
%Thus, in Fig.~\ref{fig:sg}, we have shown $U(\tilde y,\tilde t)/U_0$ for $\kappa=0.2$.
%

\section{Conclusion}

In the last decade, a disturbing trend has emerged in non-Newtonian fluid mechanics.
% with the {\it International Journal of Non-Linear Mechanics} playing a prominent role. 
Many authors have been re-deriving known results, specifically those from the 1950s through 1980s on the simple flows of certain non-Newtonian fluids. Unfortunately, many errors have been made in these ``new'' derivations. When the problems considered are actually novel, it is usually a minute change (e.g., in a boundary condition) that distinguishes them from the classical works. 
%Scores of such papers can be found in the literature of the last 15 years. 
Even when free of mathematical errors, these make little (if any) contribution to the mechanics of fluids. %; their publication is an affront to the peer-review process.

Here, we presented three known \emph{correct} solutions (by the Fourier sine transform, by the Laplace transform with the Bromwich integral inversion formula and by Laplace transform with quotient splitting and tables of inverses)  to Stokes' first problem for the second-grade and Oldroyd-B fluids. 
%These results were established 25 and 54 years ago, respectively.
Additionally, we presented a representative list of the papers in which the so-called ``new solutions'' are wrong. A complete list would be far too long to attempt here. The numerical schemes constructed in Sec.~\ref{sec:numerical} provide a wholly independent check on the classical Laplace transform solutions and the corrected Fourier sine transform solutions. 
%We hope that reviewers and editors can make use of the present work to efficiently deal with the onslaught of erroneous and incremental manuscripts with marginal scientific contributions.

Given the astonishing amount of misinformation on this topic in the literature, we offer some advice to future researchers based on the present work:
\begin{itemize}
\item The Laplace transform \emph{always} works on a well-posed linear IBVP. The solution of \citet{P84} remains the \emph{first correct exact} solution to Stokes' first problem for the second-grade fluid, despite the denigrating remarks many authors, whose work is erroneous, make about this solution. %The \emph{corrected} Fourier sine transform solution in \citep{CC10}, the alternative Laplace transform solution representation in \citep{BRG95} and the numerical solution of the IBVP all agree exactly with the solution in \citep{P84}.

\item The assumption that the fluid is initially at rest dictates that $u(y,0) \equiv 0$ (identically), whence  $\frac{\partial^k u}{\partial t^k}(y,0) \equiv 0$ for any $k$. Using the latter equality for $k\ge 1$ \emph{does not} constitute an ``additional'' or ``unphysical'' assumption as incorrectly claimed in \citep[e.g.,][]{TM05a,VNFF08} and many derivative works thereof.

\item For partial differential equations with mixed derivatives and incompatible initial and boundary data, the solution does \emph{not} have to satisfy the initial condition in backward time, i.e., as $t\to0^+$.

\item One has to be very careful in applying the Fourier sine (or cosine) transform to partial differential equations with mixed derivatives and incompatible initial and boundary data because \emph{distributional derivatives} of the Heaviside function will inevitably have to be taken for Stokes' problems. Note that the same error exposed in \citep{CJ09,CC10} is committed when solving the unsteady version of Stokes' second problem for non-Newtonian fluids. However, when $\tilde U (t) \sim t^k$ as $t\to 0$ ($k\ge1$), e.g., $\tilde U(t) = \sin(\omega t)$, the singularity is ameliorated thereby allowing an erroneous derivation to produce a correct solution. The same is true for the ``constantly accelerating plate'' problem in which $\tilde U(t) = A t$. %How to do so is typically a topic in an undergraduate course on partial differential equations, the literature would suggest many authors did not successfully complete such a course.

\item The decay boundary condition that $u\to 0$ ``sufficiently fast'' as $y\to\infty$ is typically enough to guarantee a solution by the Laplace transform. In the case of the Fourier sine transform, keeping in mind we seek a \emph{classical} solution of the partial differential equation, the implicit assumptions are made that $u$ is twice continuously differentiable in $y$, and that $u$, $\partial u/\partial y$ and $\partial^2 u/\partial y^2$ are all integrable for $y\in(0,\infty)$. (Note that this presupposes nothing about the regularity of $u$ in $t$, which we saw above is quite low.) In fact, it is known from the theory of integration \citep[see Theorem 33.7 and Proposition 23.8 in][]{p97} that these (implicit) assumptions not only guarantee that the Fourier sine transform can be applied but also that $u$ \emph{and} $\partial u/\partial y\to 0$ as $y\to\infty$. Confusion about this has led some authors to impose the additional (unnecessary) condition that $\partial u/\partial y\to 0$ as $y\to\infty$.

\item Taking the limit as the non-Newtonian parameter(s) go to zero should reduce any solution to the known Newtonian one. However, this a \emph{necessary but not sufficient} condition. Consequently, this exercise provides any insight \emph{only} if one \emph{fails} to recover the Newtonian solution, meaning the non-Newtonian one is wrong. For the problems considered in the present work, it happens that both the correct and erroneous non-Newtonian solutions reduce to the correct Newtonian one, which is unaffected by the error in computing $\frac{\partial u}{\partial t}(0,t)$ as its governing equation does not have a mixed third derivative, hence one never has to compute $\frac{\partial u}{\partial t}(0,t)$ when applying the Fourier sine transform.

\item The signs, thermodynamic restrictions on, and orders of magnitude of the non-Newtonian parameters cannot be ignored. Introducing dimensionless variables is an easy remedy to the lack of precise measurements of, e.g., the second-grade fluid's parameter $\alpha_1$. Then, plots can be made with ease and without loss of generality. %On the other hand, for the Oldroyd-B fluid, measurements of the relaxation and retardation times have been made successfully \citep{TS53}.
\end{itemize}

Another important class of non-Newtonian fluids are those of the Maxwell type, which exhibit stress relaxation but have no strain memory (retardation), i.e., $\lambda_2=0$ in Eq.~\eqref{eq:ob_const_rel}. A thorough overview of the various (correct) solution representations is given by \citet{JPB04}. And, a positivity-preserving numerical scheme, which can be used as an independent check on any suspicious ``new'' solutions that may appear in the literature, was constructed by \citet{MJ04}. Another approach to ascertaining the correctness of various solutions to Stokes' first problem is to consider the asymptotic scalings of the velocity and shear stress with time \citep{MY10} that they predict. Finally, we note that \citet{PJ87} also provide correct solutions to Stokes' first problem for viscoelastic fluids with a variety of memory kernels using the Laplace transform, while \citet{PTC88} present numerical solutions to Stokes' first problem for a class of viscoelastic fluids that reduce to the Oldroyd-B and Maxwell models in certain distinguished limits.

\section*{Acknowledgments}
The author would like to express his gratitude to Dr.\ P.\ M.\ Jordan and Prof.\ C.\ I.\ Christov for their advice and encouragement. 
%Prof.\ K.\ R.\ Rajagopal's editorial leadership in properly dealing with erroneous works in the literature is appreciated. 
This work was supported, in part, by a Walter P. Murphy Fellowship from the Robert R.\ McCormick School of Engineering and Applied Science at Northwestern University.

\bibliographystyle{model2-names}
\bibliography{stokes_1st_pb}

\begin{thebibliography}{49}
\expandafter\ifx\csname natexlab\endcsname\relax\def\natexlab#1{#1}\fi
\expandafter\ifx\csname url\endcsname\relax
  \def\url#1{\texttt{#1}}\fi
\expandafter\ifx\csname urlprefix\endcsname\relax\def\urlprefix{URL }\fi
\providecommand{\eprint}[2][]{\url{#2}}
\providecommand{\bibinfo}[2]{#2}
\ifx\xfnm\relax \def\xfnm[#1]{\unskip,\space#1}\fi
%Type = Article
\bibitem[{Amoreira and Oliveira(2010)}]{AO10}
\bibinfo{author}{Amoreira, L.J.}, \bibinfo{author}{Oliveira, P.J.},
  \bibinfo{year}{2010}.
\newblock \bibinfo{title}{Comparison of different formulations for the
  numerical calculation of unsteady incompressible viscoelastic fluid flow}.
\newblock \bibinfo{journal}{Adv. Appl. Math. Mech.} \bibinfo{volume}{4},
  \bibinfo{pages}{483--502}.
%Type = Article
\bibitem[{Amos(1969)}]{A69}
\bibinfo{author}{Amos, D.E.}, \bibinfo{year}{1969}.
\newblock \bibinfo{title}{On half-space solutions of a modified heat equation}.
\newblock \bibinfo{journal}{Quart. Appl. Math.} \bibinfo{volume}{27},
  \bibinfo{pages}{359--369}.
%Type = Article
\bibitem[{Bandelli et~al.(1995)Bandelli, Rajagopal and Galdi}]{BRG95}
\bibinfo{author}{Bandelli, R.}, \bibinfo{author}{Rajagopal, K.R.},
  \bibinfo{author}{Galdi, G.P.}, \bibinfo{year}{1995}.
\newblock \bibinfo{title}{On some unsteady motions of fluids of second grade}.
\newblock \bibinfo{journal}{Arch. Mech.} \bibinfo{volume}{47},
  \bibinfo{pages}{661--667}.
%Type = Article
\bibitem[{Christov and Jordan(2009)}]{CJ09}
\bibinfo{author}{Christov, C.I.}, \bibinfo{author}{Jordan, P.M.},
  \bibinfo{year}{2009}.
\newblock \bibinfo{title}{Comment on ``{Stokes}' first problem for an
  {Oldroyd-B} fluid in a porous half space'' {[Phys.\ Fluids 17, 023101
  (2005)]}}.
\newblock \bibinfo{journal}{Phys. Fluids} \bibinfo{volume}{21},
  \bibinfo{pages}{069101}.
%Type = Article
\bibitem[{Christov and Christov(in press)}]{CC10}
\bibinfo{author}{Christov, I.C.}, \bibinfo{author}{Christov, C.I.},
  \bibinfo{year}{in press}.
\newblock \bibinfo{title}{Comment on ``{On} a class of exact solutions of the
  equations of motion of a second grade fluid'' by {C.\ Fetec\u{a}u and J.\
  Zierep (Acta Mech.\ 150, 135--138, 2001)}}.
\newblock \bibinfo{journal}{Acta Mech.}
  \bibinfo{note}{{doi:10.1007/s00707-010-0300-2}}.
%Type = Article
\bibitem[{Coleman et~al.(1965)Coleman, Duffin and Mizel}]{CDM65}
\bibinfo{author}{Coleman, B.D.}, \bibinfo{author}{Duffin, R.J.},
  \bibinfo{author}{Mizel, V.J.}, \bibinfo{year}{1965}.
\newblock \bibinfo{title}{Instability, uniqueness, and non-existence theorems
  for the equation $u_t = u_{xx} - u_{xtx}$ on a strip}.
\newblock \bibinfo{journal}{Arch. Rational Mech. Anal.} \bibinfo{volume}{19},
  \bibinfo{pages}{100--116}.
%Type = Article
\bibitem[{Dunn and Rajagopal(1995)}]{DR95}
\bibinfo{author}{Dunn, J.E.}, \bibinfo{author}{Rajagopal, K.R.},
  \bibinfo{year}{1995}.
\newblock \bibinfo{title}{Fluids of differential type: Critical review and
  thermodynamic analysis}.
\newblock \bibinfo{journal}{Int. J. Engng Sci.} \bibinfo{volume}{33},
  \bibinfo{pages}{689--729}.
%Type = Article
\bibitem[{Erdogan(2003)}]{E03}
\bibinfo{author}{Erdogan, M.E.}, \bibinfo{year}{2003}.
\newblock \bibinfo{title}{On unsteady motions of a second-order fluid over a
  plane wall}.
\newblock \bibinfo{journal}{Int. J. Non-linear Mech.} \bibinfo{volume}{38},
  \bibinfo{pages}{1045--1051}.
%Type = Article
\bibitem[{Erdog\u{a}n and \.{I}mrak(2005)}]{EI05}
\bibinfo{author}{Erdog\u{a}n, M.E.}, \bibinfo{author}{\.{I}mrak, C.E.},
  \bibinfo{year}{2005}.
\newblock \bibinfo{title}{On unsteady unidirectional flows of a second grade
  fluid}.
\newblock \bibinfo{journal}{Int. J. Non-linear Mech.} \bibinfo{volume}{40},
  \bibinfo{pages}{1238--1251}.
%Type = Article
\bibitem[{Erdog\u{a}n and \.{I}mrak(2007a)}]{EI07a}
\bibinfo{author}{Erdog\u{a}n, M.E.}, \bibinfo{author}{\.{I}mrak, C.E.},
  \bibinfo{year}{2007}a.
\newblock \bibinfo{title}{On some unsteady flows of a non-{Newtonian} fluid}.
\newblock \bibinfo{journal}{Appl. Math. Model.} \bibinfo{volume}{31},
  \bibinfo{pages}{170--180}.
%Type = Article
\bibitem[{Erdog\u{a}n and \.{I}mrak(2007b)}]{EI07b}
\bibinfo{author}{Erdog\u{a}n, M.E.}, \bibinfo{author}{\.{I}mrak, C.E.},
  \bibinfo{year}{2007}b.
\newblock \bibinfo{title}{On the comparison of the methods used for the
  solutions of the governing equation for unsteady unidirectional flows of
  second grade fluids}.
\newblock \bibinfo{journal}{Int. J. Engng Sci.} \bibinfo{volume}{45},
  \bibinfo{pages}{786--796}.
%Type = Article
\bibitem[{Fetecau(2002)}]{F02}
\bibinfo{author}{Fetecau, C.}, \bibinfo{year}{2002}.
\newblock \bibinfo{title}{The {Rayleigh--Stokes} problem for an edge in an
  {Oldroyd-B} fluid}.
\newblock \bibinfo{journal}{C. R. Acad. Sci. Paris, Ser. I}
  \bibinfo{volume}{335}, \bibinfo{pages}{979--984}.
%Type = Article
\bibitem[{Fetecau and Fetecau(2003)}]{FF03}
\bibinfo{author}{Fetecau, C.}, \bibinfo{author}{Fetecau, C.},
  \bibinfo{year}{2003}.
\newblock \bibinfo{title}{The first problem of {Stokes} for an {Oldroyd-B}
  fluid}.
\newblock \bibinfo{journal}{Int. J. Non-Linear Mech.} \bibinfo{volume}{38},
  \bibinfo{pages}{1539--1544}.
%Type = Article
\bibitem[{Fetecau et~al.(2008)Fetecau, Hayat, Khan and Fetecau}]{FHKF08}
\bibinfo{author}{Fetecau, C.}, \bibinfo{author}{Hayat, T.},
  \bibinfo{author}{Khan, M.}, \bibinfo{author}{Fetecau, C.},
  \bibinfo{year}{2008}.
\newblock \bibinfo{title}{Unsteady flow of an {Oldroyd-B} fluid induced by the
  impulsive motion of a plate between two side walls perpendicular to the
  plate}.
\newblock \bibinfo{journal}{Acta Mech.} \bibinfo{volume}{198},
  \bibinfo{pages}{21--33}.
%Type = Article
\bibitem[{Fetecau et~al.(2009)Fetecau, Jamil, Fetecau and Vieru}]{FJFV09}
\bibinfo{author}{Fetecau, C.}, \bibinfo{author}{Jamil, M.},
  \bibinfo{author}{Fetecau, C.}, \bibinfo{author}{Vieru, D.},
  \bibinfo{year}{2009}.
\newblock \bibinfo{title}{The {Rayleigh--Stokes} problem for an edge in a
  generalized {Oldroyd-B} fluid}.
\newblock \bibinfo{journal}{Z. angew. Math. Phys. (ZAMP)} \bibinfo{volume}{60},
  \bibinfo{pages}{921--933}.
%Type = Article
\bibitem[{Fetec\v{a}u and Fetec\v{a}u(2002)}]{FF02}
\bibinfo{author}{Fetec\v{a}u, C.}, \bibinfo{author}{Fetec\v{a}u, C.},
  \bibinfo{year}{2002}.
\newblock \bibinfo{title}{The {Rayleigh--Stokes} problem for heated second
  grade fluids}.
\newblock \bibinfo{journal}{Int. J. Non-Linear Mech.} \bibinfo{volume}{37},
  \bibinfo{pages}{1011--1015}.
%Type = Book
\bibitem[{Fetter and Walecka(2003)}]{FW03}
\bibinfo{author}{Fetter, A.L.}, \bibinfo{author}{Walecka, J.D.},
  \bibinfo{year}{2003}.
\newblock \bibinfo{title}{Theoretical Mechanics of Particles and Continua}.
\newblock \bibinfo{publisher}{Dover Publications}, \bibinfo{address}{New York}.
%Type = Article
\bibitem[{Jordan(2010)}]{J10}
\bibinfo{author}{Jordan, P.M.}, \bibinfo{year}{2010}.
\newblock \bibinfo{title}{{Comments on: ``Exact solution of Stokes' first
  problem for heated generalized Burgers' fluid in a porous half-space''
  [Nonlinear Anal. RWA 9 (2008) 1628]}}.
\newblock \bibinfo{journal}{Nonlinear Anal. RWA} \bibinfo{volume}{11},
  \bibinfo{pages}{1198--1200}.
%Type = Article
\bibitem[{Jordan et~al.(2004)Jordan, Puri and Boros}]{JPB04}
\bibinfo{author}{Jordan, P.M.}, \bibinfo{author}{Puri, A.},
  \bibinfo{author}{Boros, G.}, \bibinfo{year}{2004}.
\newblock \bibinfo{title}{On a new exact solution to {Stokes}' first problem
  for {Maxwell} fluids}.
\newblock \bibinfo{journal}{Int. J. Non-linear Mech.} \bibinfo{volume}{39},
  \bibinfo{pages}{1371--1377}.
%Type = Article
\bibitem[{Jordan and Puri(2003)}]{JP03}
\bibinfo{author}{Jordan, P.M.}, \bibinfo{author}{Puri, P.},
  \bibinfo{year}{2003}.
\newblock \bibinfo{title}{{Stokes}' first problem for a {Rivlin--Ericksen}
  fluid of second grade in a porous half-space}.
\newblock \bibinfo{journal}{Int. J. Non-linear Mech.} \bibinfo{volume}{38},
  \bibinfo{pages}{1019--1025}.
%Type = Article
\bibitem[{Khan(2009)}]{K09}
\bibinfo{author}{Khan, M.}, \bibinfo{year}{2009}.
\newblock \bibinfo{title}{The {Rayleigh--Stokes} problem for an edge in a
  viscoelastic fluid with a fractional derivative model}.
\newblock \bibinfo{journal}{Nonlinear Anal. RWA} \bibinfo{volume}{10},
  \bibinfo{pages}{3190--3195}.
%Type = Article
\bibitem[{Khan et~al.(2008)Khan, Alia, Hayat and Fetecau}]{KAHF08}
\bibinfo{author}{Khan, M.}, \bibinfo{author}{Alia, S.H.},
  \bibinfo{author}{Hayat, T.}, \bibinfo{author}{Fetecau, C.},
  \bibinfo{year}{2008}.
\newblock \bibinfo{title}{{MHD} flows of a second grade fluid between two side
  walls perpendicular to a plate through a porous medium}.
\newblock \bibinfo{journal}{Int. J. Non-linear Mech.} \bibinfo{volume}{43},
  \bibinfo{pages}{302--319}.
%Type = Article
\bibitem[{Khan et~al.(2010)Khan, Malik, Fetecau and Fetecau}]{KMFF10}
\bibinfo{author}{Khan, M.}, \bibinfo{author}{Malik, R.},
  \bibinfo{author}{Fetecau, C.}, \bibinfo{author}{Fetecau, C.},
  \bibinfo{year}{2010}.
\newblock \bibinfo{title}{Exact solutions for the unsteady flow of a {Burgers}Õ
  fluid between two sidewalls perpendicular to the plate}.
\newblock \bibinfo{journal}{Chem. Eng. Comm.} \bibinfo{volume}{197},
  \bibinfo{pages}{1367--1386}.
%Type = Book
\bibitem[{Kolmogorov and Fomin(1975)}]{KF75}
\bibinfo{author}{Kolmogorov, A.N.}, \bibinfo{author}{Fomin, S.V.},
  \bibinfo{year}{1975}.
\newblock \bibinfo{title}{Introductory Real Analysis}.
\newblock \bibinfo{publisher}{Dover Publications}, \bibinfo{address}{New York}.
%Type = Article
\bibitem[{Mickens and Jordan(2004)}]{MJ04}
\bibinfo{author}{Mickens, R.E.}, \bibinfo{author}{Jordan, P.M.},
  \bibinfo{year}{2004}.
\newblock \bibinfo{title}{A positivity-preserving nonstandard finite difference
  scheme for the damped wave equation}.
\newblock \bibinfo{journal}{Numer. Methods Partial Differential Eq.}
  \bibinfo{volume}{20}, \bibinfo{pages}{639--649}.
%Type = Article
\bibitem[{Morrison(1956)}]{M56}
\bibinfo{author}{Morrison, J.A.}, \bibinfo{year}{1956}.
\newblock \bibinfo{title}{Wave propagation in rods of {Voigt} material and
  visco-elastic materials with three-parameter models}.
\newblock \bibinfo{journal}{Quart. Appl. Math.} \bibinfo{volume}{14},
  \bibinfo{pages}{153--169}.
%Type = Article
\bibitem[{Muzychka and Yovanovich(2010)}]{MY10}
\bibinfo{author}{Muzychka, Y.S.}, \bibinfo{author}{Yovanovich, M.M.},
  \bibinfo{year}{2010}.
\newblock \bibinfo{title}{Unsteady viscous flows and {Stokes}'s first problem}.
\newblock \bibinfo{journal}{Int. J. Thermal Sci.} \bibinfo{volume}{49},
  \bibinfo{pages}{820--828}.
%Type = Article
\bibitem[{Oldroyd(1950)}]{O50}
\bibinfo{author}{Oldroyd, J.G.}, \bibinfo{year}{1950}.
\newblock \bibinfo{title}{On the formulation of rheological equations of
  state}.
\newblock \bibinfo{journal}{Proc. R. Soc. Lond. A} \bibinfo{volume}{200},
  \bibinfo{pages}{523--541}.
%Type = Article
\bibitem[{Oldroyd(1958)}]{O58}
\bibinfo{author}{Oldroyd, J.G.}, \bibinfo{year}{1958}.
\newblock \bibinfo{title}{Non-{Newtonian} effects in steady motion of some
  idealized elastico-viscous liquids}.
\newblock \bibinfo{journal}{Proc. R. Soc. Lond. A} \bibinfo{volume}{245},
  \bibinfo{pages}{278--297}.
%Type = Article
\bibitem[{{Phan-Thien} and Chew(1988)}]{PTC88}
\bibinfo{author}{{Phan-Thien}, N.}, \bibinfo{author}{Chew, Y.T.},
  \bibinfo{year}{1988}.
\newblock \bibinfo{title}{On the {Rayleigh} problem for a viscoelastic fluid}.
\newblock \bibinfo{journal}{J. Non-Newtonian Fluid Mech.} \bibinfo{volume}{28},
  \bibinfo{pages}{117--127}.
%Type = Article
\bibitem[{Preziosi and Joseph(1987)}]{PJ87}
\bibinfo{author}{Preziosi, L.}, \bibinfo{author}{Joseph, D.D.},
  \bibinfo{year}{1987}.
\newblock \bibinfo{title}{{Stokes}' first problem for viscoelastic fluids}.
\newblock \bibinfo{journal}{J. Non-Newtonian Fluid Mech.} \bibinfo{volume}{25},
  \bibinfo{pages}{239--259}.
%Type = Book
\bibitem[{Priestley(1997)}]{p97}
\bibinfo{author}{Priestley, H.A.}, \bibinfo{year}{1997}.
\newblock \bibinfo{title}{Introduction to Integration}.
\newblock \bibinfo{publisher}{Oxford University Press},
  \bibinfo{address}{Oxford}.
%Type = Article
\bibitem[{Puri(1984)}]{P84}
\bibinfo{author}{Puri, P.}, \bibinfo{year}{1984}.
\newblock \bibinfo{title}{Impulsive motion of a flat plate in a
  {Rivlin--Ericksen} fluid}.
\newblock \bibinfo{journal}{Rheol. Acta} \bibinfo{volume}{23},
  \bibinfo{pages}{451--453}.
%Type = Article
\bibitem[{Rajagopal and Srinivasa(2000)}]{RS00}
\bibinfo{author}{Rajagopal, K.R.}, \bibinfo{author}{Srinivasa, A.R.},
  \bibinfo{year}{2000}.
\newblock \bibinfo{title}{A thermodynamic frame work for rate type fluid
  models}.
\newblock \bibinfo{journal}{J. Non-Newtonian Fluid Mech.} \bibinfo{volume}{88},
  \bibinfo{pages}{207--227}.
%Type = Book
\bibitem[{Schlichting(1979)}]{s79}
\bibinfo{author}{Schlichting, H.}, \bibinfo{year}{1979}.
\newblock \bibinfo{title}{Boundary-Layer Theory}.
\newblock \bibinfo{publisher}{McGraw-Hill}, \bibinfo{address}{New York}.
  \bibinfo{edition}{7th} edition.
%Type = Article
\bibitem[{Shen et~al.(2006)Shen, Tan, Zhao and Masuoka}]{STZM06}
\bibinfo{author}{Shen, F.}, \bibinfo{author}{Tan, W.}, \bibinfo{author}{Zhao,
  Y.}, \bibinfo{author}{Masuoka, T.}, \bibinfo{year}{2006}.
\newblock \bibinfo{title}{The {Rayleigh--Stokes} problem for a heated
  generalized second grade fluid with fractional derivative model}.
\newblock \bibinfo{journal}{Nonlinear Anal. RWA} \bibinfo{volume}{7},
  \bibinfo{pages}{1072--1080}.
%Type = Article
\bibitem[{Stokes(1851)}]{S51}
\bibinfo{author}{Stokes, G.G.}, \bibinfo{year}{1851}.
\newblock \bibinfo{title}{On the effect of the internal friction of fluids on
  the motion of pendulums}.
\newblock \bibinfo{journal}{Trans. Cambridge Phil. Soc.} \bibinfo{volume}{9
  (Part II)}, \bibinfo{pages}{8--106}.
%Type = Book
\bibitem[{Strikwerda(2004)}]{S04}
\bibinfo{author}{Strikwerda, J.}, \bibinfo{year}{2004}.
\newblock \bibinfo{title}{Finite Difference Schemes and Partial Differential
  Equations}.
\newblock \bibinfo{publisher}{SIAM}, \bibinfo{address}{Philadelphia}.
%Type = Article
\bibitem[{Tan and Masuoka(2005a)}]{TM05b}
\bibinfo{author}{Tan, W.}, \bibinfo{author}{Masuoka, T.},
  \bibinfo{year}{2005}a.
\newblock \bibinfo{title}{{Stokes}' first problem for a second grade fluid in a
  porous half-space with heated boundary}.
\newblock \bibinfo{journal}{Int. J. Non-Linear Mech.} \bibinfo{volume}{40},
  \bibinfo{pages}{515--522}.
%Type = Article
\bibitem[{Tan and Masuoka(2005b)}]{TM05a}
\bibinfo{author}{Tan, W.}, \bibinfo{author}{Masuoka, T.},
  \bibinfo{year}{2005}b.
\newblock \bibinfo{title}{{Stokes}' first problem for an {Oldroyd-B} fluid in a
  porous half space}.
\newblock \bibinfo{journal}{Phys. Fluids} \bibinfo{volume}{17},
  \bibinfo{pages}{023101}.
%Type = Article
\bibitem[{Tanner(1962)}]{T62}
\bibinfo{author}{Tanner, R.I.}, \bibinfo{year}{1962}.
\newblock \bibinfo{title}{Note on the {Rayleigh} problem for a visco-elastic
  fluid}.
\newblock \bibinfo{journal}{Z. angew. Math. Phys. (ZAMP)} \bibinfo{volume}{13},
  \bibinfo{pages}{573--580}.
%Type = Article
\bibitem[{Ting(1963)}]{T63}
\bibinfo{author}{Ting, T.W.}, \bibinfo{year}{1963}.
\newblock \bibinfo{title}{Certain non-steady flows of second-order fluids}.
\newblock \bibinfo{journal}{Arch. Rational Mech. Anal.} \bibinfo{volume}{14},
  \bibinfo{pages}{1--26}.
%Type = Article
\bibitem[{Toms and Strawbridge(1953)}]{TS53}
\bibinfo{author}{Toms, B.A.}, \bibinfo{author}{Strawbridge, D.J.},
  \bibinfo{year}{1953}.
\newblock \bibinfo{title}{Elastic and viscous properties of dilute solutions of
  polymethyl methacrylate in organic liquids}.
\newblock \bibinfo{journal}{Trans. Faraday Soc.} \bibinfo{volume}{49},
  \bibinfo{pages}{1225--1232}.
%Type = Article
\bibitem[{Townsend(1973)}]{T73}
\bibinfo{author}{Townsend, P.}, \bibinfo{year}{1973}.
\newblock \bibinfo{title}{Numerical solutions of some unsteady flows of
  elastico-viscous liquids}.
\newblock \bibinfo{journal}{Rheol. Acta} \bibinfo{volume}{12},
  \bibinfo{pages}{13--18}.
%Type = Book
\bibitem[{Truesdell and Rajagopal(2000)}]{TR00}
\bibinfo{author}{Truesdell, C.}, \bibinfo{author}{Rajagopal, K.R.},
  \bibinfo{year}{2000}.
\newblock \bibinfo{title}{An Introduction to the Mechanics of Fluids}.
\newblock \bibinfo{publisher}{Birkh\"auser}, \bibinfo{address}{Boston}.
%Type = Article
\bibitem[{Vieru et~al.(2008)Vieru, Nazar, Fetecau and Fetecau}]{VNFF08}
\bibinfo{author}{Vieru, D.}, \bibinfo{author}{Nazar, M.},
  \bibinfo{author}{Fetecau, C.}, \bibinfo{author}{Fetecau, C.},
  \bibinfo{year}{2008}.
\newblock \bibinfo{title}{New exact solutions corresponding to the first
  problem of {Stokes} for {Oldroyd-B} fluids}.
\newblock \bibinfo{journal}{Comput. Math. Appl.} \bibinfo{volume}{55},
  \bibinfo{pages}{1644--1652}.
%Type = Article
\bibitem[{Zierep and Bohning(2008)}]{ZB08}
\bibinfo{author}{Zierep, J.}, \bibinfo{author}{Bohning, R.},
  \bibinfo{year}{2008}.
\newblock \bibinfo{title}{Conservation of energy of non-{Newtonian} media for
  the {Rayleigh--Stokes} problem}.
\newblock \bibinfo{journal}{Acta Mech.} \bibinfo{volume}{201},
  \bibinfo{pages}{5--11}.
%Type = Article
\bibitem[{Zierep et~al.(2007)Zierep, Bohning and Fetecau}]{ZBF07}
\bibinfo{author}{Zierep, J.}, \bibinfo{author}{Bohning, R.},
  \bibinfo{author}{Fetecau, C.}, \bibinfo{year}{2007}.
\newblock \bibinfo{title}{{Rayleigh--Stokes} problem for non-{Newtonian} medium
  with memory}.
\newblock \bibinfo{journal}{Z. Angew. Math. Mech. (ZAMM)} \bibinfo{volume}{87},
  \bibinfo{pages}{462--467}.
%Type = Article
\bibitem[{Zierep and Fetecau(2007)}]{ZF07}
\bibinfo{author}{Zierep, J.}, \bibinfo{author}{Fetecau, C.},
  \bibinfo{year}{2007}.
\newblock \bibinfo{title}{Energetic balance for the {Rayleigh--Stokes} problem
  of a second grade fluid}.
\newblock \bibinfo{journal}{Int. J. Engng Sci.} \bibinfo{volume}{45},
  \bibinfo{pages}{155--162}.

\end{thebibliography}

\end{document}